%% file: main.tex
\documentclass{article}
\usepackage{spconf,graphicx,hyperref}
\usepackage{placeins} 
\usepackage{amsmath,amssymb}
\usepackage{bm}
\usepackage{subcaption}
\usepackage{svg}
\usepackage{tabularx}
\usepackage{booktabs}
\usepackage{multirow}
\usepackage{makecell}
\usepackage{cite} 

\title{TISDiSS: A Training-Time and Inference-Time Scalable Framework for Discriminative Source Separation}
\name{
    \begin{tabular}{c}
		Yongsheng Feng$^{1}$, Yuetonghui Xu$^{1}$, Jiehui Luo$^{1}$, Hongjia Liu$^{1}$, Xiaobing Li$^{1}$, Feng Yu$^{1}$, Wei Li $^{2,3\sthanks{Corresonding Author}}$
    \end{tabular}
}
\address{
	$^1$ Department of Music AI and Music IT, Central Conservatory of Music, Beijing, China \\
  $^2$ School of Computer Science and Technology, Fudan University, Shanghai, China \\
  $^3$ Shanghai Key Laboratory of Intelligent Information Processing, Fudan University, Shanghai, China\\
}

\begin{document}

\ninept
\maketitle
\begin{abstract}
  Source separation is a fundamental task in speech, music, and audio processing, and it also provides cleaner and larger data for training generative models. 
  However, improving separation performance in practice often depends on increasingly large networks, inflating training and deployment costs. 
  Motivated by recent advances in inference-time scaling for generative modeling, we propose \textbf{T}raining-Time and \textbf{I}nference-Time \textbf{S}calable \textbf{Di}scriminative \textbf{S}ource \textbf{S}eparation (TISDiSS), 
  a unified framework that integrates early-split multi-loss supervision, shared-parameter design, and dynamic inference repetitions.
  TISDiSS enables flexible speed–performance trade-offs by adjusting inference depth without retraining additional models. 
  We further provide systematic analyses of architectural and training choices and show that training with more inference repetitions improves shallow-inference performance, benefiting low-latency applications. 
  Experiments on standard speech separation benchmarks demonstrate state-of-the-art performance with a reduced parameter count, establishing TISDiSS as a scalable and practical framework for adaptive source separation. 
  Code is available at \url{https://github.com/WingSingFung/TISDiSS}.
\end{abstract}

\begin{keywords}
	speech separation, source separation, discriminative models, inference-time scaling, training-time scaling
\end{keywords}

\FloatBarrier


\input{sections/Introduction}

\input{sections/Experiments/frameworks}

\input{sections/Methods}
\vspace{-4pt}
\input{sections/Experiment}
\vspace{-4pt}
\input{sections/Conclusion}

\newpage
\bibliographystyle{IEEEbib}
\bibliography{strings,refs}

\end{document}

%% file: sections/Introduction.tex
\section{Introduction}
\label{sec:intro}

Source separation is a fundamental problem in speech, music, and general audio processing. It not only supports end applications such as real-time communication, hearing aids, and voice assistants, but also enables the creation of cleaner and larger datasets that benefit downstream generative tasks including text-to-speech, text-to-music, and audio synthesis \cite{xuTIGER2025,heEmilia2024,yuAutoprep2024,baiSeedMusic2024}. In practice, user requirements can vary widely: devices with limited computational resources demand faster inference to obtain usable separated audio, while high-performance systems may favor separation quality regardless of inference costs. 

However, achieving stronger separation performance usually relies on training deeper and wider models \cite{saijoTFLocoformer2024,shinSepReformer2024,luBS-Roformer2023,wangTFGridNet2023}, which requires extensive computational resources, making training and deployment expensive.
Meanwhile, recent advances in large-scale generative models have shown an \emph{inference-time scaling} phenomenon: increasing inference iterations can improve output quality without changing model parameters \cite{openaiO12024,guoDeepseekR12025,yeLlasa2025,kangLLaSEG12025}. For discriminative source separation, this phenomenon brings forward a key direction: designing a single model that scales performance at inference time to reduce the need for training multiple large models.

To address this, we propose \textbf{T}raining-Time and \textbf{I}nference-Time \textbf{S}calable \textbf{Di}scriminative \textbf{S}ource \textbf{S}eparation (TISDiSS), the first framework unifying:
\vspace{-4pt}
\begin{itemize}
    \item \emph{early-split multi-loss supervision}, which constrains intermediate representations and improves the effectiveness of early-split separation models \cite{leeSepTDA2024,shinSepReformer2024};
    \item \emph{shared-parameter design}, which reduces model size for lightweight deployment \cite{luoTinySepformer2022,ohPapez2024,xuTIGER2025};
    \item \emph{dynamic inference repetitions}, which enable flexible speed--performance trade-offs by adjusting computational depth during inference.
\end{itemize}
\vspace{-4pt}

Unlike prior work, TISDiSS leverages these techniques jointly to realize inference-time scalability with a single trained model. We further conduct systematic analyses of early-split supervision, multi-loss settings, shared-parameter design, and model structure, providing insights into their roles and interactions. In addition, we introduce a simple training strategy: training with more inference repetitions consistently improves shallow-inference performance, offering a practical solution for low-latency separation.
To validate the framework, we focus on speech separation for its well-established benchmarks and efficient experimental comparisons. 
Experiments on WSJ0-2mix \cite{hersheyWSJ02mix2015}, Libri2Mix \cite{cosentinoLibriMixOpenSourceDataset2020}, and WHAMR! \cite{maciejewskiWHAMR2020} demonstrate that TISDiSS achieves state-of-the-art(SOTA) performance while supporting both training-time and inference-time scalability. 

%% file: sections/Experiments/frameworks.tex
\begin{figure*}[t!] 
    \centering
    \begin{minipage}[t]{\textwidth}
        \centering
        \includegraphics[width=0.95\textwidth, keepaspectratio]{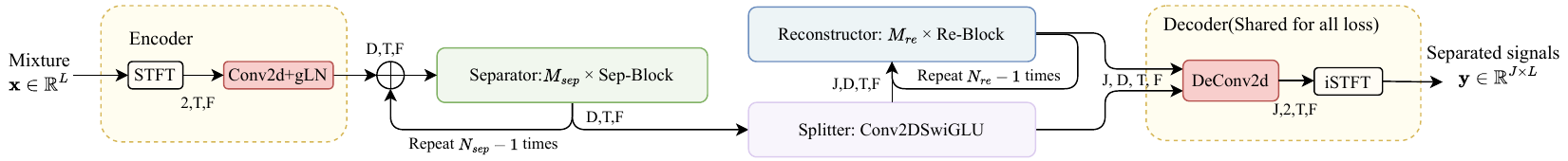}
        \subcaption{Overall architecture}
        \label{fig:framework-main}
    \end{minipage}
    \vspace{2pt}
    \begin{minipage}[t]{0.43\textwidth}
        \centering
        \includegraphics[width=\textwidth, keepaspectratio]{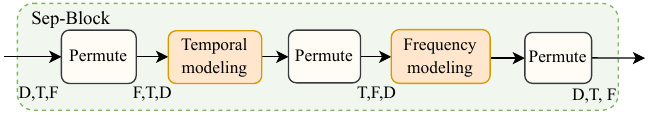}
        \subcaption{Separation block}
        \label{fig:sepblock}
    \end{minipage}
    \hfill
    \begin{minipage}[t]{0.5\textwidth}
        \centering
        \includegraphics[width=\textwidth, keepaspectratio]{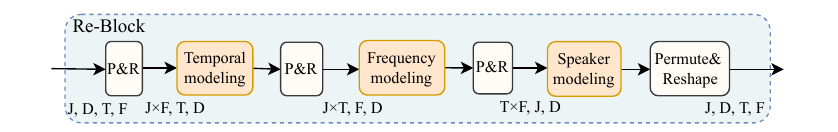}
        \subcaption{Reconstruction block (P\&R: Permute\&Reshape)}
        \label{fig:reblock}
    \end{minipage}
    
    \vspace{-5pt}
    \caption{Framework of the proposed TISDiSS method. (a) Overall architecture, (b) separation block, and (c) reconstruction block.}
    \label{fig:framework}
    \vspace{-14pt}
\end{figure*}

%% file: sections/Methods.tex
\section{Method}
\label{sec:method}
\subsection{Framework Overview}
\autoref{fig:framework-main} presents the proposed TISDiSS framework, designed primarily for Time-Frequency (TF)-domain models. The mono mixed signal $\bm{x} \in \mathbb{R}^{L}$ is generated by superposing $J$ speech signals $\bm{s} \in \mathbb{R}^{J \times L}$ and one noise signal $\bm{b} \in \mathbb{R}^{L}$, with $\bm{x} = \sum\nolimits_{j=1}^J \bm{s}_{j} + \bm{b}$. Here, $L$ denotes the number of time-domain samples, and $j=1,\dots,J$ indexes speech sources. TISDiSS comprises five core components: Encoder, Separator, Splitter, Reconstructor, and Decoder, whose detailed implementations are described as follows.

\textbf{Encoder:} The input $\bm{x}$ is processed in two main steps. First, the Short-Time Fourier Transform (STFT) converts $\bm{x}$ into TF-domain features $\bm{X} \in \mathbb{R}^{2 \times T \times F}$, where $T$ is the number of time frames, $F$ is the STFT frequency bin count, and the dimension 2 corresponds to the real and imaginary parts of the spectrum. Second, these features are processed through a Conv2D layer followed by global layer normalization (gLN) to generate the final feature $\bm{Z} \in \mathbb{R}^{D \times T \times F}$, following the operation $\bm{Z} = \text{gLN}\left(\text{Conv2D}(\bm{X})\right)$ where $D$ denotes the dimension of the output feature channels.

\textbf{Separator:} The Encoder’s output $\bm{Z}$ is processed by a shared-parameter Separator for $N_{\text{sep}}$ iterations. Each Separator contains $M_{\text{sep}}$ Sep-Blocks (structure in \autoref{fig:sepblock}), which are dual-path blocks referencing TF-locoformer \cite{saijoTFLocoformer2024} to capture feature information in the time ($T$) and frequency ($F$) dimensions. In this work, the temporal modeling and frequency modeling modules of the Separator exclusively adopt the SOTA TF-locoformer as the base model, ensuring consistency and fairness for subsequent comparative experiments.

\textbf{Splitter:} The Separator’s output is decomposed into $J$ speaker-specific features $\bm{V} \in \mathbb{R}^{J \times D \times T \times F}$ via a Splitter. The core of the Splitter is a Conv2DSwiGLU module, improved from the Conv1DSwiGLU in TF-locoformer \cite{saijoTFLocoformer2024}; this improvement enables simultaneous analysis of feature information in both the time ($T$) and frequency ($F$) dimensions.

\textbf{Reconstructor:} The Splitter’s output $\bm{V}$ is refined by a shared-parameter Reconstructor for $N_{\text{re}}$ iterations. Each Reconstructor contains $M_{\text{re}}$ Re-Blocks (structure in \autoref{fig:reblock}), which are triple-path modules referencing SepTDA \cite{leeSepTDA2024} to capture feature information in time ($T$), frequency ($F$), and speaker ($J$) dimensions. Unlike SepTDA (which uses a simple Transformer for the speaker dimension $J$), this work employs the same module structure for $J$ as for $T$ and $F$, reducing variable interference and simplifying the fairness of comparative experiments.

\textbf{Decoder:} Outputs of the Splitter and Reconstructor are converted into time-domain target signals $\bm{y} \in \mathbb{R}^{J \times L}$: first, a DeConv2D scheme maps features back to an STFT complex spectrum $\bm{Y} \in \mathbb{R}^{J \times 2 \times T \times F}$ with dimensions aligned to the input TF-domain features; finally, the inverse STFT (iSTFT) recovers the time-domain signal to yield clean speaker speech $\bm{y}$.

\vspace{-4pt}
\subsection{Training Objective}
TISDiSS adopts a multi-loss supervision mechanism to constrain model intermediate stages, facilitating scalable TF-domain representation learning. The loss design includes: Final Output Loss, which uses permutation-invariant scale-invariant signal-to-noise ratio (SI-SNR) loss for the Decoder’s target signal (denoted $L_{\text{last}}$); and Intermediate Auxiliary Losses, which introduce SI-SNR auxiliary losses at intermediate outputs of the Separator, Splitter, and Reconstructor (denoted $L_{\text{sep}}$, $L_{\text{split}}$, $L_{\text{re}}$, respectively).

It should be noted that the calculation of $L_{\text{split}}$ adopts the output of the last Separator, which results in the equivalence between $L_{\text{sep}}$ corresponding to the last Separator and $L_{\text{split}}$. Thus, $L_{\text{sep}}$ averages only intermediate outputs of the first $N_{\text{sep}}-1$ Separators, and $L_{\text{re}}$ averages intermediate outputs of the first $N_{\text{re}}-1$ Reconstructors. The overall training loss is a weighted average of all activated loss terms, with the mathematical expression shown in Eq. \ref{eq:total_loss}:

\begin{equation}
\label{eq:total_loss}
L = \frac{1}{K} \left[ 
\lambda_{\text{last}} L_{\text{last}} + 
\lambda_{\text{sep}} \overline{L}_{\text{sep}} + 
\lambda_{\text{split}} L_{\text{split}} + 
\lambda_{\text{re}} \overline{L}_{\text{re}}
\right]
\end{equation}

where $K$ denotes the total number of activated loss terms (e.g., $K=1$ for only $L_{\text{last}}$, $K=3$ for $L_{\text{last}}+L_{\text{re}}+L_{\text{split}}$), $\lambda_{\text{last}}$, $\lambda_{\text{sep}}$, $\lambda_{\text{split}}$, $\lambda_{\text{re}}$ are weights of respective loss terms, $\overline{L}_{\text{sep}} = \frac{1}{N_{\text{sep}}-1}\sum_{i=1}^{N_{\text{sep}}-1} L_{\text{sep},i}$ and $\overline{L}_{\text{re}} = \frac{1}{N_{\text{re}}-1}\sum_{i=1}^{N_{\text{re}}-1} L_{\text{re},i}$. For baseline comparison ease, all activated loss terms use a weight of 1; 
subsequent ablation experiments verify the effectiveness of different loss term selections, and identify $L_{\text{last}}$, $L_{\text{re}}$, and $L_{\text{split}}$ as the optimal loss configuration for model training.

\vspace{-4pt}
\subsection{Inference-Time Scaling}
TISDiSS’s core advantage is flexible scalability during inference: a single set of trained model weights achieves a performance-efficiency trade-off by adjusting the parameter pair $(N_{\text{sep}}, N_{\text{re}})$ (number of Separator/Reconstructor iterations). Reducing $N_{\text{sep}}$ and $N_{\text{re}}$ lowers inference latency for limited computational resources, while increasing them enhances model representation capability for higher separation quality.

For further performance optimization, short-term fine-tuning can be performed on existing weights after increasing $N_{\text{sep}}$ and $N_{\text{re}}$ (no training from scratch), avoiding redundant costs of training dedicated models for different application scenarios.

%% file: sections/Experiment.tex
\section{EXPERIMENTS}
\label{sec:exper}
\vspace{-4pt}
\subsection{Dataset and Experimental Setup}
We evaluate TISDiSS on three standard speech separation corpora: WSJ0-2mix \cite{hersheyWSJ02mix2015}, Libri2Mix \cite{cosentinoLibriMixOpenSourceDataset2020}, and WHAMR! \cite{maciejewskiWHAMR2020}. All experiments use the fully overlapped ``min'' version of the data with a unified sampling rate of 8 kHz. Specifically, the durations of the train/val/test splits for WSJ0-2mix and Libri2Mix are approximately 30/10/5 hours and 212/11/11 hours, respectively; WHAMR! is the noisy and reverberant variant of WSJ0-2mix.

Model implementation is based on the ESPnet-SE framework \cite{liESPnetSE2021}.To ensure fair comparison, the parameter settings for the Encoder, Decoder, and the modules responsible for modeling the time ($T$), frequency ($F$), and speaker ($J$) dimensions within Sep-Blocks and Re-Blocks in this study all adhere to the same configurations as the medium-sized setting of TF-locoformer \cite{saijoTFLocoformer2024}. 
The naming convention for TISDiSS models is: TISDiSS-sep\(\{M_{\text{sep}}\}\times\{N_{\text{sep}}\}\)-re\(\{M_{\text{re}}\}\times\{N_{\text{re}}\}\) (\(N_{\text{re}}\) value used during inference)-l\(\{\text{loss configuration}\}\). If no explicit parentheses are included, the \(N_{\text{sep}}\) and \(N_{\text{re}}\) values used during training are adopted for inference.

The AdamW optimizer is employed for training, with a weight decay coefficient of \(1\times10^{-2}\). The learning rate is linearly warmed up to \(1\times10^{-3}\) over the first 2,000 steps; if the validation loss fails to improve for 3 consecutive epochs, the learning rate is halved. 
Training is capped at 150 epochs, with early stopping triggered if the validation loss fails to improve for 10 consecutive epochs. The final reported model is obtained by performing parameter averaging on the 5 checkpoint models with the lowest validation loss. During the fine-tuning phase, the only parameter modification is setting the learning rate to start from $1\times10^{-4}$.

The experiments adopt SI-SNR improvement (SI-SNRi) and signal-to-distortion ratio improvement (SDRi) \cite{vincentPerformanceMeasurementBlind2006} as primary evaluation metrics.

\input{sections/Experiments/wsj0-2mix.tex}
\input{sections/Experiments/libri2mix+whamr.tex}

\vspace{-4pt}
\subsection{Comparison with SOTA Models}
\label{sec:sota-comparison}
\autoref{tab:wsj0-2mix} presents experimental results on the WSJ0-2mix dataset, where TISDiSS is compared with two SOTA models: SepReformer (a time-domain model) and TF-locoformer (a TF-domain model). 
Notably, TISDiSS-sep\(1\!\times\!2\)-re\(1\!\times\!6\)(8) does not employ the dynamic mixing strategy—yet it still achieves higher SI-SNRi and SDRi than the Large versions of the two aforementioned models (which do adopt dynamic mixing). Additionally, TISDiSS requires significantly fewer parameters than these Large models, highlighting its efficiency advantage. 

\autoref{tab:whamrlibri2mix} presents experimental results on the Libri2Mix and WHAMR! datasets. Specifically, results on WHAMR! (a noise-reverberation corrupted corpus) and Libri2Mix (a larger-scale corpus) collectively demonstrate TISDiSS’s capability to stably enhance separation performance across both noisy-reverberant scenarios and large-scale data.
Across different TISDiSS configurations, consistent performance gains of approximately 1 dB in both SI-SNRi and SDRi are observed compared to prior SOTA models.

\input{sections/Experiments/comparison.tex}

\input{sections/Experiments/8ablation-study.tex}

\vspace{-4pt}
\subsection{Ablation Study: Effects of Early-Split, Multi-Loss, and Shared-Parameter}
\label{subsec:ablation}

\autoref{tab:comparison} evaluates how early-split, multi-loss, and shared-parameter configurations affect model performance on WSJ0-2mix.

First, TF-locoformer(M)-sep$6\!\times\!1$ refers to results from the original TF-locoformer paper, while TF-locoformer(M)-sep$6\!\times\!1$(R) is our reproduction under the same training environment for fair comparison. Under non-early-split settings, adding multi-loss supervision directly to the original architecture (TF-locoformer-sep$6\!\times\!1$-l$6\!\times\!1$) degrades performance—indicating naive multi-loss application on undivided features is unbeneficial.

Next, we analyzed shared parameters’ impact on non-early-split models. To match computational complexity, we set $M_{\text{sep}}=3, 2, 1$ with corresponding $N_{\text{sep}}=2, 3, 6$. Results show shared parameters cause performance loss (consistent with prior work\cite{luoTinySepformer2022,ohPapez2024}), and this loss diminishes as $M_{\text{sep}}$ increases.

To improve shared-parameter models, we optimized TF-loco-former via residual connections (preserving original features to reduce learning difficulty\cite{xuTIGER2025}) and a Decoder-preceding Splitter. The optimized model (TISDiSS-sep$6\!\times\!1$-l$6\!\times\!1$) outperforms its TF-locoformer counterpart; even with shared parameters, its performance loss is far smaller than the TF-locoformer baseline—validating these optimizations.

Under early-split settings, we assessed the effect of multi-loss supervision using TISDiSS-sep$2\!\times\!1$-re$3\!\times\!1$-l1 as the single-loss baseline (trained with $L_{\text{last}}$ only). We compared four loss configurations: $L_{\text{last}}$ (baseline, ``-l1''), $L_{\text{last}}+L_{\text{re}}$ (``-l3''), $L_{\text{last}}+L_{\text{re}}+L_{\text{split}}$ (``-l1+3''), and $L_{\text{last}}+L_{\text{re}}+L_{\text{split}}+L_{\text{sep}}$ (``-l1$\times$2+3''). All multi-loss variants outperform the baseline, with ``-l3'' ($L_{\text{last}}+L_{\text{re}}$) achieving the best results—aligning with SepReformer\cite{shinSepReformer2024} findings that direct Reconstructor supervision drives iterative performance gains.

Finally, we studied the shared-parameter early-split model TISDiSS-sep\(1\!\times\!2\)-re\(1\!\times\!3\). Comparing its four variants with non-shared-parameter models reveals that the performance losses induced by shared parameters span 0.05 dB to 0.5 dB in magnitude—all of which are consistently small, well within the range of negligible differences for practical purposes. When $N_{\text{sep}}$/$N_{\text{re}}$ are consistent between inference and training, the four variants show no meaningful performance difference. Thus, \autoref{fig:loss} compares their scaling performance under varying inference-time $N_{\text{re}}$. Among them, ``-l1+3'' ($L_{\text{last}}+L_{\text{re}}+L_{\text{split}}$) exhibits the most stable scaling across Reconstructor repetitions—underscoring its superior inference-time scaling capability. This setup is therefore adopted as the default loss configuration for all other experiments.

\vspace{-4pt}
\subsection{Ablation Study: Training- and Inference-Time Scalability}
\label{subsec:ablation-train-infer}
\vspace{-3pt}
This subsection presents ablation experiments—primarily conducted on the WSJ0-2mix dataset—that examine how training configurations (e.g., $N_{\text{sep}}$, $N_{\text{re}}$, $M_{\text{re}}$, and shared-parameter / multi-loss design) affect inference-time scalability and separation performance.

\autoref{fig:en_comparison} compares results for $N_{\text{sep}}=1$ and $N_{\text{sep}}=2$: under identical conditions, increasing $N_{\text{sep}}$ improves overall performance.

\autoref{fig:re_comparison} shows performance differences across $N_{\text{re}}=2, 3, 4, 6$. Increasing $N_{\text{re}}$ during training boosts inference performance without added parameters; critically, models trained with larger $N_{\text{re}}$ outperform those trained with smaller $N_{\text{re}}$ even when using smaller $N_{\text{re}}$ at inference. For example, the model trained with $N_{\text{re}}=4$ achieves higher SI-SNRi at inference $N_{\text{re}}=2$ and $3$ than counterparts trained with $N_{\text{re}}=2$ or $3$—guiding lightweight TISDiSS training.

\autoref{fig:Mre_Nre_comparison} compares two configurations: $M_{\text{re}}=1, N_{\text{re}}=4$ and $M_{\text{re}}=2, N_{\text{re}}=2$ (same-color boxes denote equal inference cost). The former (increasing $N_{\text{re}}$) outperforms the latter (increasing $M_{\text{re}}$), confirming that under fixed inference cost, increasing $N_{\text{re}}$ yields more significant gains.

\autoref{fig:re_comparison} and \autoref{fig:Mre_Nre_comparison} also show that small inference $N_{\text{re}}$ harms scaling (even causing degradation). However, TISDiSS’s flexible shared-parameter and multi-loss architecture mitigates this via fine-tuning with larger $N_{\text{re}}$. As \autoref{fig:wsj0-2mix-finetune} shows, fine-tuning en$1\!\times\!2$-re$1\!\times\!3$-l1+1x3 (training $N_{\text{re}}=3$) to en$1\!\times\!2$-re$1\!\times\!3$-l1+1x6 (training $N_{\text{re}}=6$) improves performance at larger inference $N_{\text{re}}$.

\autoref{fig:whamr-finetune} and \autoref{fig:librimix-finetune} further demonstrate this with en$1\!\times\!2$-re$2\!\times\!2$-l1+1x2 (trained with $N_{\text{re}}=2$), which exhibits ``feature hallucination'' and degraded scaling at inference $N_{\text{re}}\geq4$ due to insufficient training. Fine-tuning with $N_{\text{re}}=4$ (yielding en$1\!\times\!2$-re$2\!\times\!4$-l1+1x4) effectively restores and enhances inference scalability.

\vspace{-4pt}
\subsection{Ablation Study: Lightweight Configuration Variants}
\label{subsec:lightweight-ablation}
\autoref{fig:other_comparison} evaluates the performance of TISDiSS under various lightweight configurations on the WSJ0-2mix dataset. The TF-domain model sep1$\times$2-re1$\times$3-l1+1$\times$3 serves as the baseline, chosen to demonstrate the feasibility of the separation framework in supporting lightweight design explorations.

sep1$\times$2-re1$\times$3-l1+1$\times$3-spksplitconv2d replaces the baseline’s Conv2dSwiGLU splitter with a simpler Conv2d module. Compared to Conv2dSwiGLU, Conv2d has significantly lower computational complexity and fewer parameters, with only a minor performance drop—making it a preferable splitter choice for lightweight model requirements.

Existing studies show that introducing band-split significantly reduces memory usage and improves performance on 16 kHz speech datasets and the 44.1 kHz MUSDB18HQ dataset\cite{xuTIGER2025, luoBS-RNN2023, luBS-Roformer2023, wangMelBandRoFormer2023}, but its effect on 8 kHz speech datasets remains untested. Inspired by TIGER\cite{xuTIGER2025} and MelFormer\cite{wangMelBandRoFormer2023}, we adapted TISDiSS by replacing the Conv2D in the Encoder and DeConv2D in the Decoder with TIGER’s band-split module and band-restoration module, respectively. 
Specifically, with an STFT window size $N=128$, the frequency dimension $\lfloor N/2 \rfloor + 1 = 65$ was compressed to 33 (adopting the band-split pattern from TIGER) and 32 (using the band-split scheme from MelFormer) bins, aiming to further reduce memory consumption and accelerate computation.
Results for sep1$\times$2-re1$\times$3-l1+1$\times$3-bandsplit and sep1$\times$2-re1$\times$3-l1+1$\times$3-melbandsplit show that band-split operations lead to minor performance degradation. This is likely because 8 kHz sampling only covers up to 4 kHz, making information loss from frequency compression more pronounced than in 16 kHz and 44.1 kHz data.

%% file: sections/Experiments/wsj0-2mix.tex
\begin{table}[t]
  \centering
  \setlength{\tabcolsep}{0.8mm}{

  \centering
  \caption{Comparisons with prior methods on WSJ0-2mix with and without dynamic mixing (DM).
  Results in [dB]. 
  \vspace{-7pt}
  }
  
  \label{tab:wsj0-2mix}
  
  \begin{tabularx}{\columnwidth}{Xccc}
    \toprule
    Methods                                   & Param [M] & SI-SNRi       & SDRi                \\ \midrule
    SepReformer-B\cite{shinSepReformer2024}                             & 14.2      & 23.8          & 23.9                \\
    SepReformer-L+DM\cite{shinSepReformer2024}                          & 59.4      & 25.1          & 25.2                \\ \midrule
    TF-Locoformer-M\cite{saijoTFLocoformer2024}                           & 15.0      & 23.6          & 23.8                \\
    TF-Locoformer-M+DM\cite{saijoTFLocoformer2024}                        & 15.0      & 24.6          & 24.7                \\
    TF-Locoformer-L\cite{saijoTFLocoformer2024}                           & 22.5      & 24.2          & 24.3                \\
    TF-Locoformer-L+DM\cite{saijoTFLocoformer2024}                        & 22.5      & 25.1          & 25.2                \\ \midrule
    TISDiSS-sep$1\!\times\!2$-re$1\!\times\!3$ (3)  & 8.0      & 23.9          & 24.0                \\
    TISDiSS-sep$1\!\times\!2$-re$1\!\times\!3$ (5)  & 8.0      & 24.3          & 24.4                \\ \midrule
    TISDiSS-sep$1\!\times\!2$-re$1\!\times\!6$ (3)  & 8.0      & 24.4          & 24.5               \\
    TISDiSS-sep$1\!\times\!2$-re$1\!\times\!6$ (6)  & 8.0      & 25.1          & 25.2                \\
    TISDiSS-sep$1\!\times\!2$-re$1\!\times\!6$ (8)  & 8.0      & \textbf{25.2} & \textbf{25.3}       \\ \bottomrule
  \end{tabularx}
  }
\vspace{-10pt}
\end{table}

%% file: sections/Experiments/libri2mix+whamr.tex
\begin{table}[t]
  \centering
  \setlength{\tabcolsep}{0.2mm}{

  \centering
  \caption{Comparisons with prior methods on WHAMR! and Libri2Mix.
  Results in [dB]. 
  \vspace{-7pt}
  }
  
  \label{tab:whamrlibri2mix}

  \begin{tabularx}{\columnwidth}{Xccc}
  \toprule
  \multirow{2}*{\textbf{Methods}} & \multirow{2}*{\makecell[c]{\textbf{Param} \\ \textbf{[M]}} } & \textbf{WHAMR!} & \textbf{Libri2Mix} \\
  \cmidrule(lr){3-3} \cmidrule(lr){4-4} & & \makecell[c]{SI-SNRi/SDRi} & \makecell[c]{SI-SNRi/SDRi} \\
  \midrule
  
  TF-GridNet\cite{wangTFGridNet2023} & 14.4 & 17.1/15.6 & -/- \\
  SepReformer-L + DM & 59.4 & 17.1/16.0 & -/- \\
  TF-Locoformer-S & 5.0 & 17.4/15.9 & -/- \\
  TF-Locoformer-M & 15.0 & 18.5/16.9 & 22.1/22.2 \\
  FLA-TFLocoformer-M\cite{wangFLASepformer2025} & 15.1 & 18.7/17.0 & 22.2/22.4 \\
  \midrule
  TISDiSS-sep$1\!\times\!2$-re$1\!\times\!3$ (3) & 8.0 & 19.6/17.9 & 23.0/23.2 \\
  TISDiSS-sep$1\!\times\!2$-re$1\!\times\!3$ (4) & 8.0 & 19.8/18.1 & 23.1/23.3 \\
  TISDiSS-sep$1\!\times\!2$-re$2\!\times\!2$ (2) & 11.2 & 19.8/18.1 & 23.3/23.6 \\
  TISDiSS-sep$1\!\times\!2$-re$2\!\times\!2$ (3) & 11.2 & \textbf{19.9}/\textbf{18.2} & \textbf{23.5}/\textbf{23.7} \\
  \bottomrule
  
  \end{tabularx}
  }
\vspace{-12pt}
\end{table}

%% file: sections/Experiments/comparison.tex
\begin{table}[t]
  \centering
  \setlength{\tabcolsep}{0.1mm}{

  \centering
  \caption{Ablation study on WSJ0-2mix — effects of early-split (ES), shared-parameter design (SP), and multi-loss supervision (ML). Results in [dB].
  \vspace{-7pt}
  }

  \label{tab:comparison}
  
  \begin{tabularx}{\columnwidth}{X c c c c}
    \toprule
    Methods & ES/SP/ML & Param [M] & SI-SNRi & SDRi \\
    \midrule
    TF-locoformer(M)-sep$6\!\times\!1$\cite{saijoTFLocoformer2024} &
      $\times$/$\times$/$\times$ & 15.0 & 23.64 & 23.78 \\
    TF-locoformer(M)-sep$6\!\times\!1$(R) &
      $\times$/$\times$/$\times$ & 15.0 & 23.31 & 23.45 \\
    \midrule
    TF-locoformer-sep$6\!\times\!1$-l$6\!\times\!1$ &
      $\times$/$\times$/\checkmark & 15.0 & 23.02 & 23.16 \\
    TF-locoformer-sep$3\!\times\!2$-l$1\!\times\!2$ &
      $\times$/\checkmark/\checkmark & 7.5 & 22.26 & 22.41 \\
    TF-locoformer-sep$2\!\times\!3$-l$1\!\times\!3$ &
      $\times$/\checkmark/\checkmark & 5.0 & 21.82 & 21.96 \\
    TF-locoformer-sep$1\!\times\!6$-l$1\!\times\!6$ &
      $\times$/\checkmark/\checkmark & 2.5 & 20.88 & 21.05 \\
    
    \midrule
    TISDiSS-sep$6\!\times\!1$-l$6\!\times\!1$ &
      $\times$/$\times$/\checkmark & 17.3 & 23.33 & 23.32 \\
    TISDiSS-sep$2\!\times\!3$-l$1\!\times\!3$ &
      $\times$/\checkmark/\checkmark & 7.3 & 22.77 & 22.91 \\
    TISDiSS-sep$1\!\times\!6$-l$1\!\times\!6$ &
      $\times$/\checkmark/\checkmark & 4.8 & 22.16 & 22.30 \\
    \midrule
    TISDiSS-sep$2\!\times\!1$-re$3\!\times\!1$-l$1$ &
      \checkmark/$\times$/$\times$ & 16.8 & 24.00 & 24.13 \\
    TISDiSS-sep$2\!\times\!1$-re$3\!\times\!1$-l$3$ &
      \checkmark/$\times$/\checkmark & 16.8 & 24.44 & 24.57 \\
    TISDiSS-sep$2\!\times\!1$-re$3\!\times\!1$-l$1\!+\!3$ &
      \checkmark/$\times$/\checkmark & 16.8 & 24.04 & 24.17 \\
    TISDiSS-sep$2\!\times\!1$-re$3\!\times\!1$-l$1\!\times\!2\!+\!3$ &
      \checkmark/$\times$/\checkmark & 16.8 & 24.29 & 24.42 \\
    \midrule
    TISDiSS-sep$1\!\times\!2$-re$1\!\times\!3$-l$1$ &
      \checkmark/\checkmark/$\times$ & 8.0 & 23.95 & 24.08 \\
    TISDiSS-sep$1\!\times\!2$-re$1\!\times\!3$-l$3$ &
      \checkmark/\checkmark/\checkmark & 8.0 & 23.92 & 24.05 \\
    TISDiSS-sep$1\!\times\!2$-re$1\!\times\!3$-l$1\!+\!3$ &
      \checkmark/\checkmark/\checkmark & 8.0 & 23.94 & 24.08 \\
    TISDiSS-sep$1\!\times\!2$-re$1\!\times\!3$-l$1\!\times\!2\!+\!3$ &
      \checkmark/\checkmark/\checkmark & 8.0 & 23.89 & 24.02 \\
    \bottomrule
  \end{tabularx}
  }
\vspace{-12pt}
\end{table}

%% file: sections/Experiments/8ablation-study.tex
\begin{figure*}[t]
    \centering
    \begin{subfigure}[t]{0.24\textwidth}
      \centering
      \includegraphics[width=\linewidth]{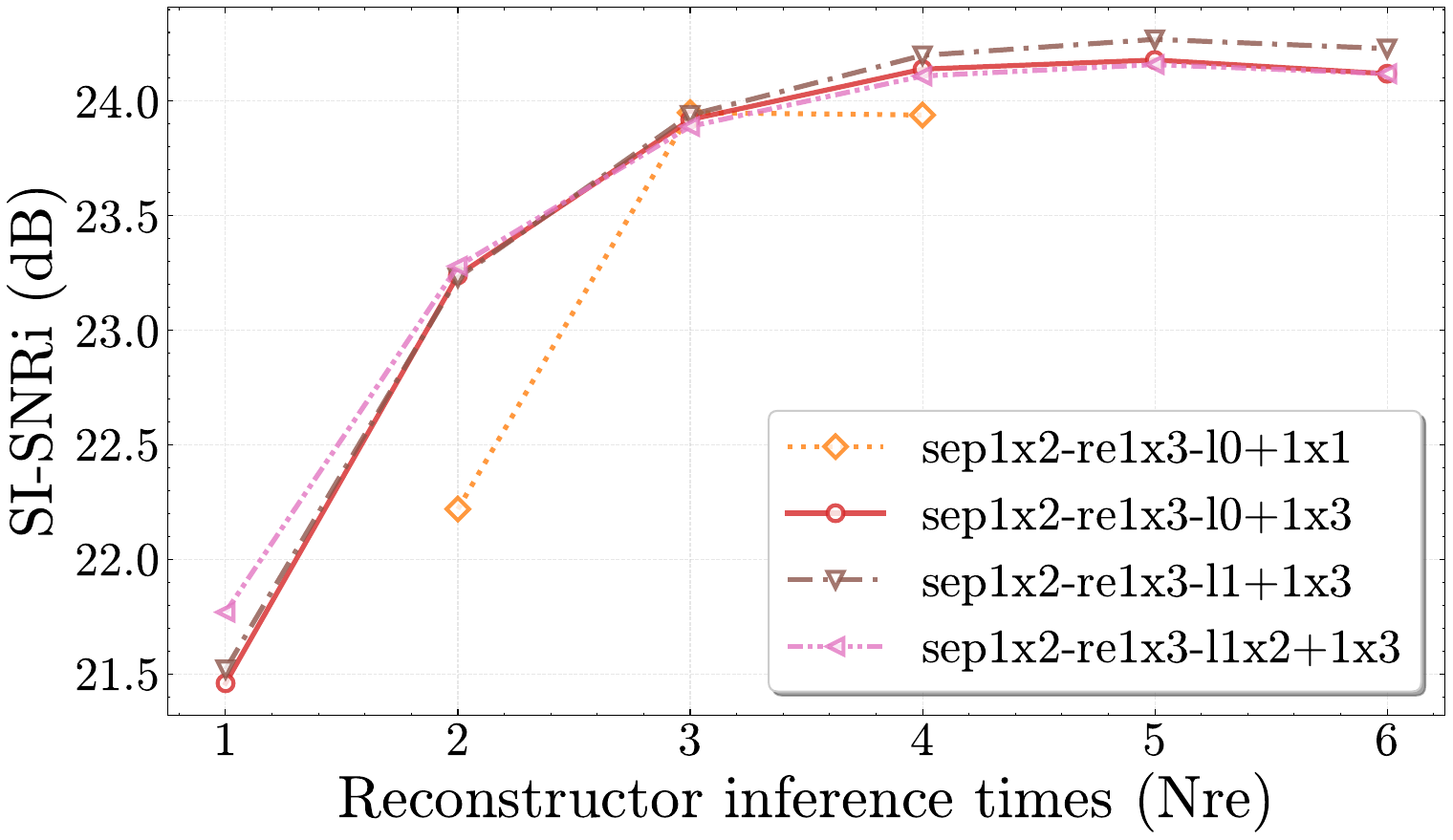}
      \caption{Comparison of training losses}
      \label{fig:loss}
    \end{subfigure}\hfill
    \begin{subfigure}[t]{0.24\textwidth}
      \centering
      \includegraphics[width=\linewidth]{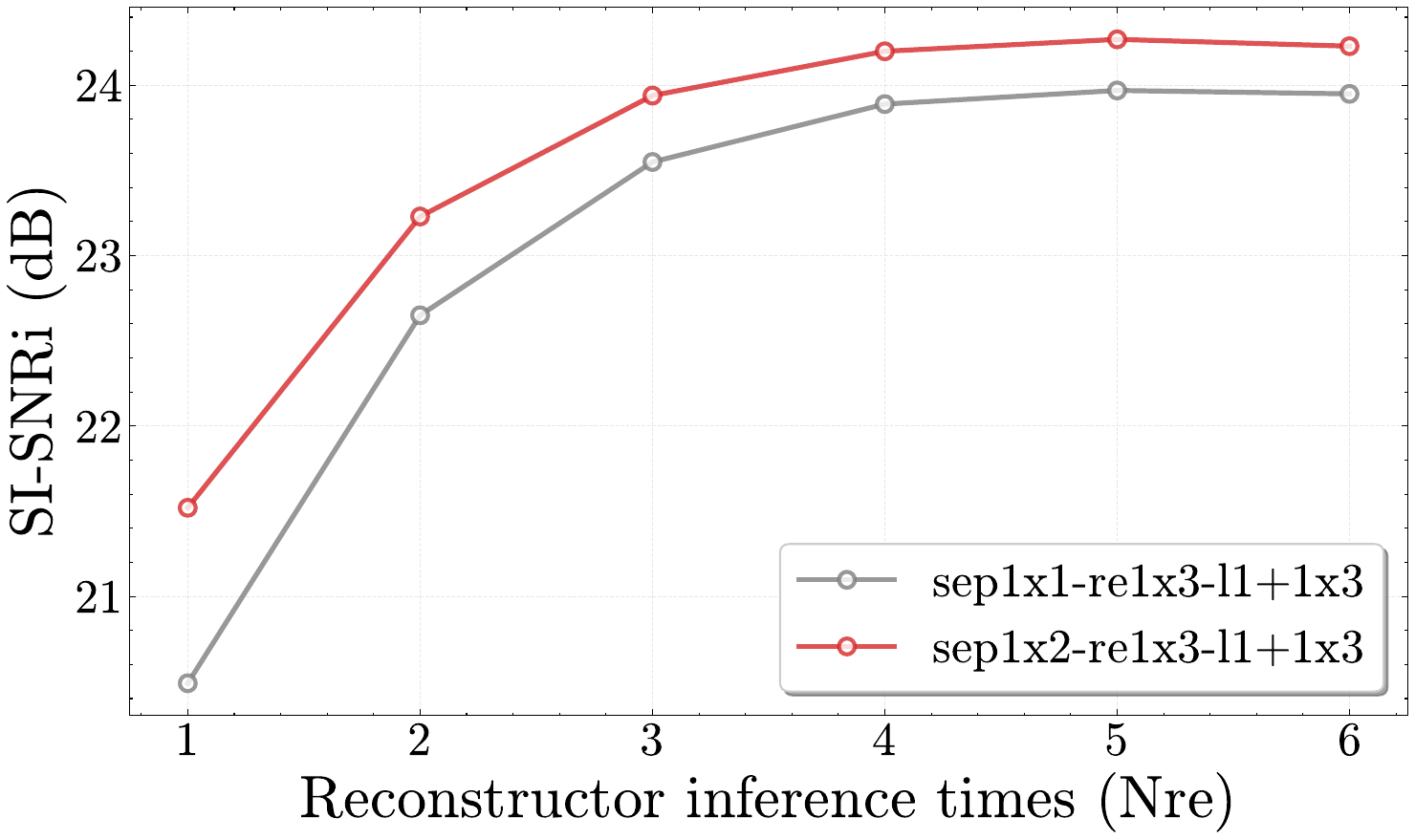}
      \caption{Comparison of $N_{\text{sep}}$}
      \label{fig:en_comparison}
    \end{subfigure}\hfill
    \begin{subfigure}[t]{0.24\textwidth}
      \centering
      \includegraphics[width=\linewidth]{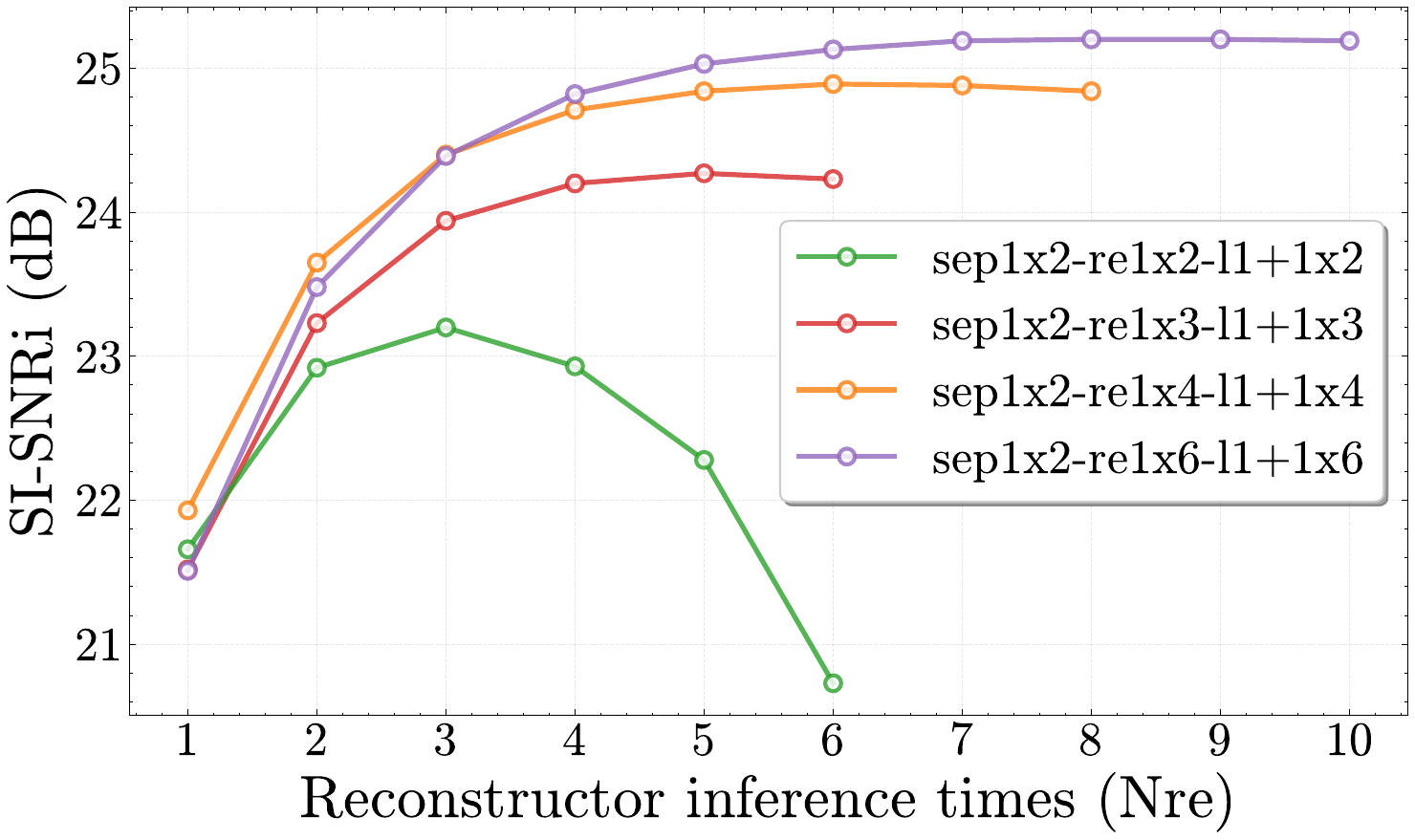}
      \caption{Comparison of $N_{\text{re}}$}
      \label{fig:re_comparison}
    \end{subfigure}\hfill
    \begin{subfigure}[t]{0.24\textwidth}
      \centering
      \includegraphics[width=\linewidth]{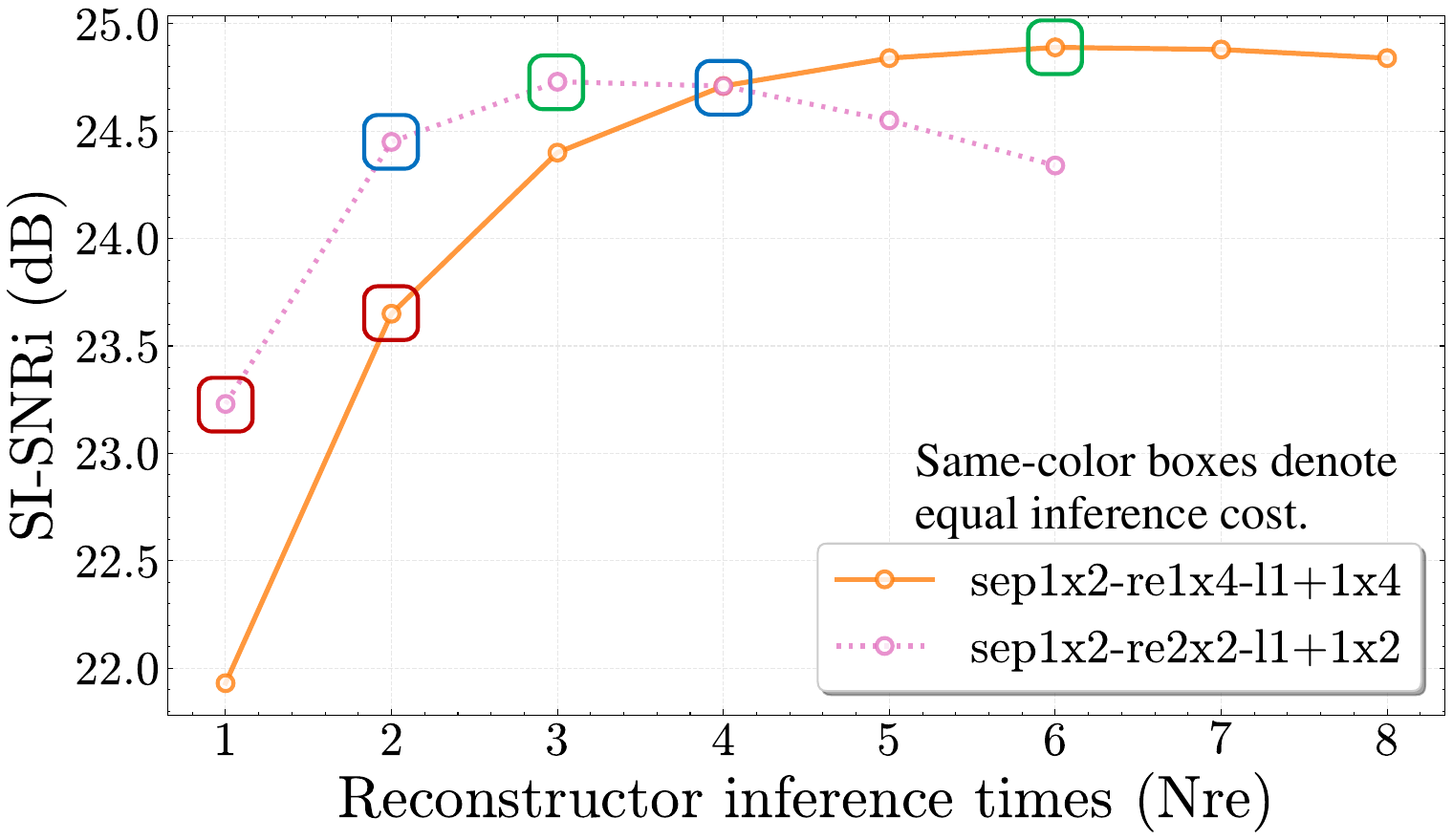}
      \caption{Comparison of $M_{\text{re}}$ and $N_{\text{re}}$}
      \label{fig:Mre_Nre_comparison}
    \end{subfigure}
    \vspace{2pt}
    \begin{subfigure}[t]{0.24\textwidth}
      \centering
      \includegraphics[width=\linewidth]{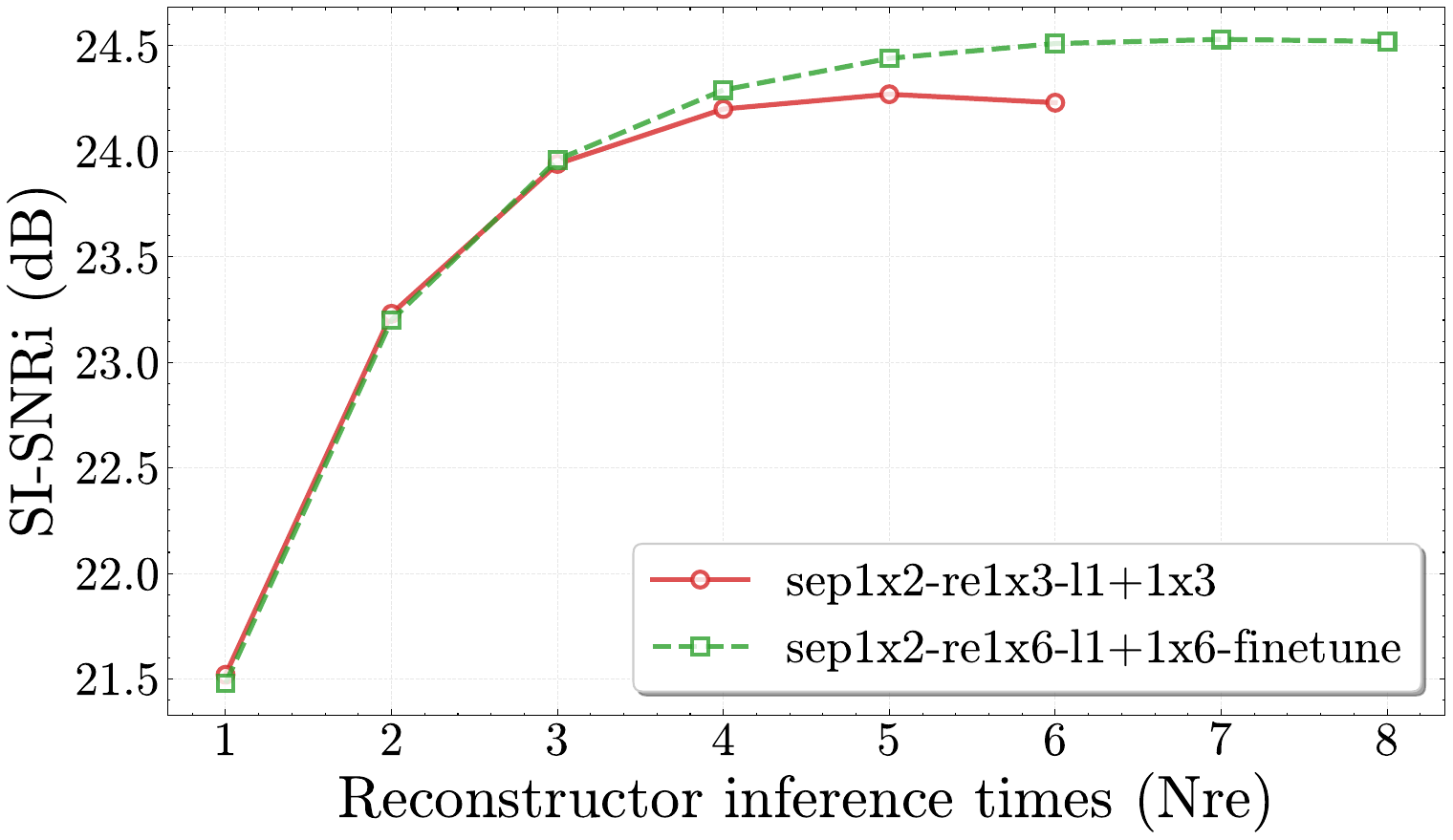}
      \caption{Fine-tuning on WSJ0-2mix}
      \label{fig:wsj0-2mix-finetune}
    \end{subfigure}\hfill
    \begin{subfigure}[t]{0.24\textwidth}
      \centering
      \includegraphics[width=\linewidth]{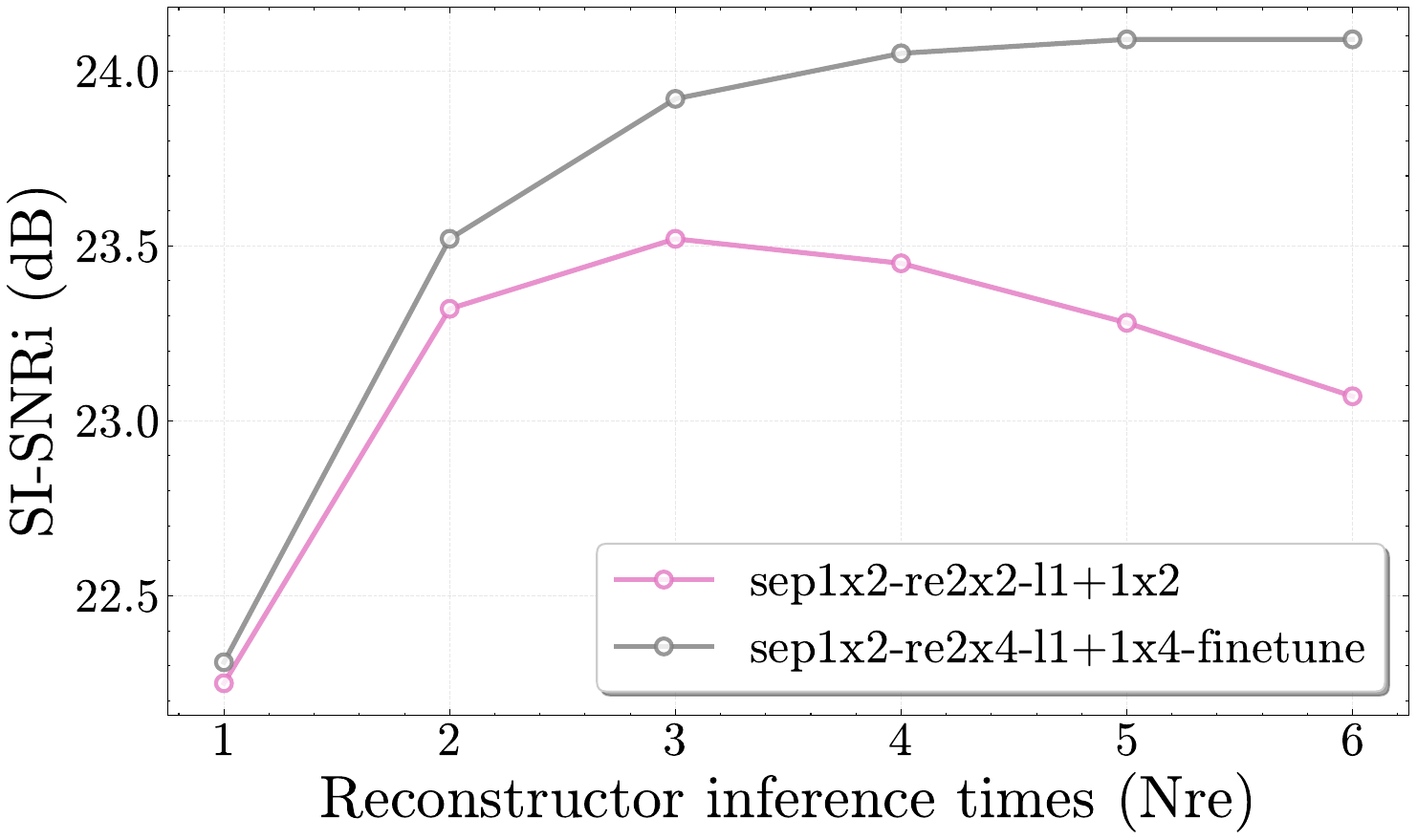}
      \caption{Fine-tuning on Libri2Mix}
      \label{fig:librimix-finetune}
    \end{subfigure}\hfill
    \begin{subfigure}[t]{0.24\textwidth}
      \centering
      \includegraphics[width=\linewidth]{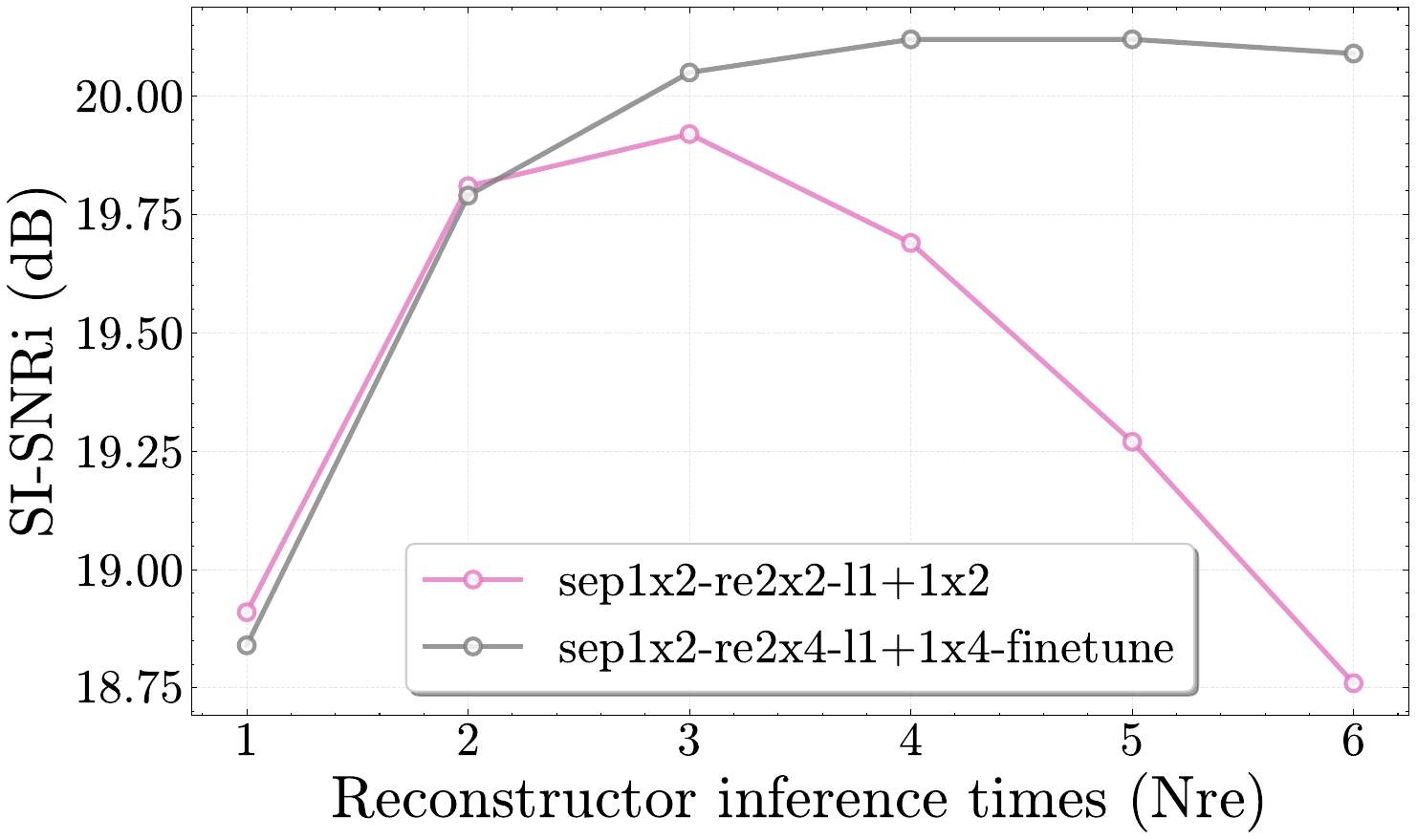}
      \caption{Fine-tuning on WHAMR!}
      \label{fig:whamr-finetune}
    \end{subfigure}\hfill
    \begin{subfigure}[t]{0.24\textwidth}
      \centering
      \includegraphics[width=\linewidth]{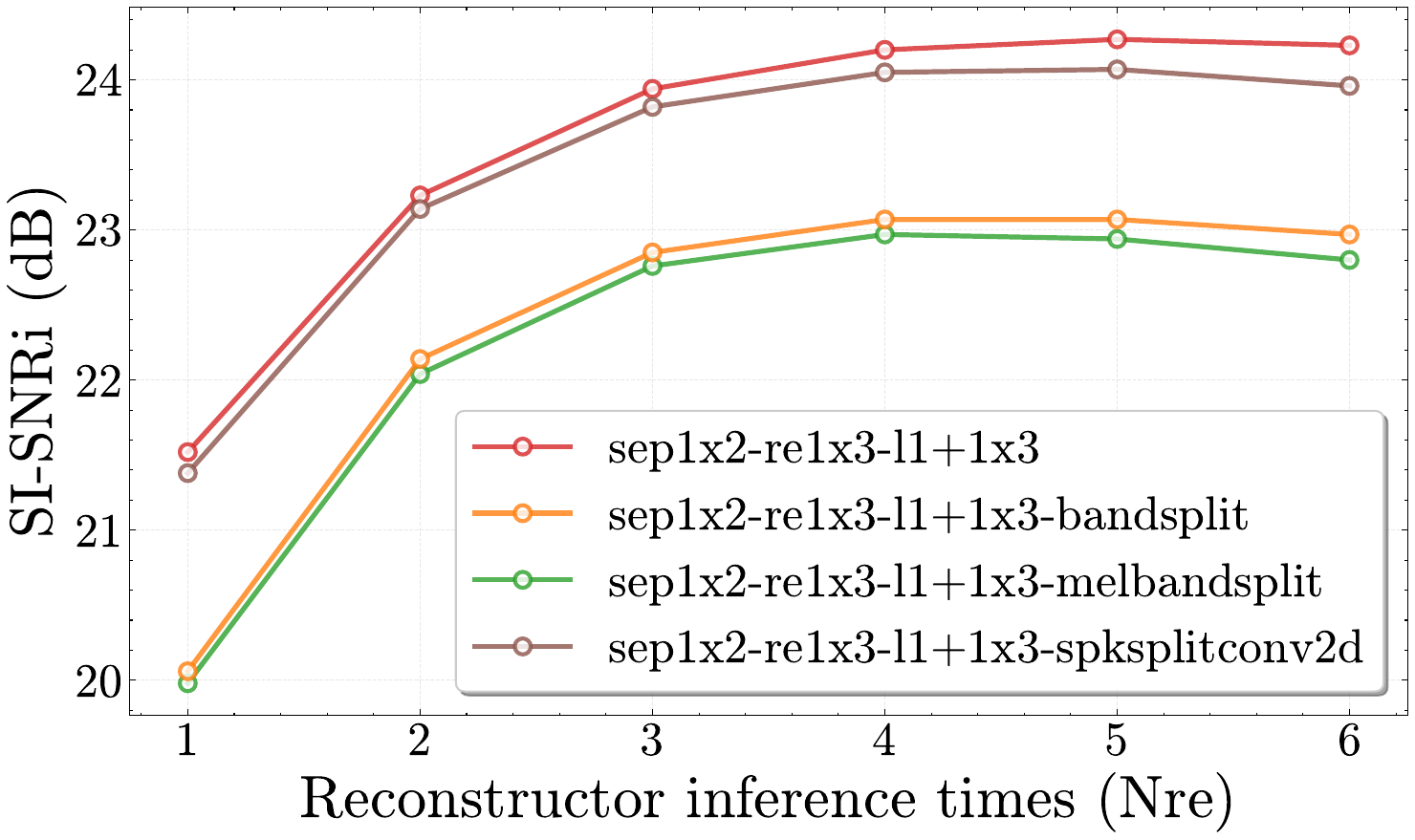}
      \caption{Lightweight configuration variants}
      \label{fig:other_comparison}
    \end{subfigure}\hfill
    \vspace{-5pt}
    \caption{Ablation study with SI-SNRi [dB] on the $y$-axis and $N_{\text{re}}$ used during inference on the $x$-axis (WSJ0-2mix except (f)(g)).}
    \label{fig:ablation-study}
    \vspace{-14pt}
\end{figure*}

%% file: sections/Conclusion.tex
\section{Conclusion}
We presented TISDiSS, which achieves both training-time and inference-time scalability for source separation. 
It unifies early-split multi-loss supervision, shared-parameter design, and dynamic inference repetitions, enabling flexible speed–performance trade-offs and improving shallow-inference performance through deeper training. 
Experiments on standard speech separation benchmarks demonstrate state-of-the-art results with fewer parameters, establishing TISDiSS as a practical paradigm for adaptive audio processing.